\documentstyle[tighten,amsfonts,aps]{revtex}
\begin{document}
\draft
\title{Verification of the semiclassical method for an electron
\protect \\ moving in a homogeneous magnetic field}
\author{V. A. Bordovitsyn\thanks{E-mail: bord@mail.tomsknet.ru} and
A. N. Myagkii\thanks{E-mail: myagkii@mail.ru}}
\address{Physics Department, Tomsk State University, Tomsk 634050, Russia}
\date{\today}
\maketitle
\begin{abstract}
A procedure based on the semiclassical approximation for high energy
levels is developed to yield solutions to the classical equation of
charge motion and to the Bargmann-Michel-Telegdi spin equation.
To this end, exact solutions to the Klein-Gordon and
the Dirac-Pauli equations are used. The essence of the procedure under
review is that the quantum state of a charged particle in a homogeneous
magnetic field is represented as a superposition of states corresponding
to the neighboring energy levels. As a consequence, the behavior of the
expectation values of the momentum and spin operators with respect to
the resulting nonstationary wave function (packet) strictly obey the
classical equations of charge motion and spin precession.
\end{abstract} 

%\narrowtext
\section{Introduction}

In the semiclassical approach, the charge motion is described in a
classical way, and quantum transitions are calculated on the base
of matrix elements~\cite{4}. This method is known to be fairly in the
ultrarelativistic case at ultrahigh electron energies. For example,
at energies of $\sim$ 2.5 GeV, which are typical for current storage
rings with a characteristic parameter $H\rho\sim 10^7$ Oe cm
the electron energy levels amount to $n\sim 10^{17}$. For radiative
transitions to lower energy levels, $\Delta n/n\sim 10^{-7}$. In this case,
according to the uncertainty principle, all quantum processes occur in
a region as small as a few angstroms (see \cite{1} for details).
Clearly, this is a negligible small value for an orbital radius of
about 10 m. Hence, the classical, or more precisely, semiclassical concept
of the electron trajectory can be used here.

In the theory of synchrotron radiation, the semiclassical method allows
all quantum corrections to the synchrotron radiation power, including the
effects with spin-flip to be calculated~\cite{2,3}. This method is
straightforward to use. What is more, it provides a visual picture of
physical processes occurring at ultrahigh electron energy (recoil effects,
radiative spin self-polarization, spin magnetic-moment radiation, mixed
charge and magneton radiation, radiation associated with the anomalous part
of electron magnetic moment, etc.).

Despite of the impressive success of the semiclassical synchrotron-radiation
theory rigorous mathematical substantiation of this method based on
exact solutions to the Klein-Gordon equation or to the Dirac-Pauli equation
is, to our knowledge, lacking heretofore. The basic statements of the theory
were, in fact, postulated on the basis of the uncertainty principle
(see~\cite{4}).

In this paper, solutions to the classical equation of charge motion and to
the Bargmann-Michel-Telegdi (BMT) spin precession equation~\cite{bmt} are
constructed in the framework of quantum mechanics, using nonstationary
wave functions (see general statements in \cite{fg}). The later are, in
their turn, exact solutions to the Klein-Gordon equation or to the
Dirac-Pauli equation for a charged particle moving in a homogeneous
magnetic field.  A simplified version of this method was used
in~\cite{5} to reveal a physical pattern underlying the behavior of
longitudinal electron-spin polarization in a magnetic field.  This
approach is adopted in~\cite{2} to describe quantum neutron spin-flip
transitions.

Our task is to explore the time evolution of the expectation
values of the momentum and spin operators for a charged particle moving
in a homogeneous magnetic field. To demonstrate the capability of
the method discussed, let us examine the problem at hand first.
In Sec.\ \ref{sec2}, we will consider a scalar particle and compare
the time evolution of the expectation value of the momentum operator for
this particle with the solution to the classical equation of charge
motion. These are found to differ by a factor.
Sec.\ \ref{sec3} will focus on a spin-$\frac{1}{2}$ particle.
In addition to evolution of the expectation value of the
momentum operator, we will discuss the behavior of the
expectation value of the spin operator. The latter also differs from
the solution to the BMT equation by factor only. Finally,
in Sec.\ \ref{sec4}, we will introduce a procedure that will help to
eliminate the difference between the quantum and classical approaches.
To this end, the state of a particle will be represented by a wave
packet so that the time evolution of expectation values of the
corresponding operators with respect to this nonstationary wave
function will coincide with that predicted classical mechanics.

\section{Motion of a scalar particle}
\label{sec2}

Let us assume that state of scalar particle is formed as a wave packet
by superposing exact solutions to the Klein-Gordon equation
corresponding to three neighboring energy levels $n-1$, $n$, and $n+1$:
$$\Psi({\bf r},t)=\sum_{m=n-1}^{n+1}A_{m} \psi_{m}({\bf
r})\exp(-\frac{i}{\hbar}m_0c^2B_{m}t),$$ where $\psi_{m}({\bf r})$ are
the stationary solutions to the Klein-Gordon equation in a homogeneous
magnetic field (see Appendix\ \ref{appendixA}) and $A_{m}$ are the
expansion coefficients.

To determine the coefficients $A_{m}$ in the semiclassical approximation
($n\gg 1$), the probability that a particle will be found at each
of the levels is assumed to be the same. For $n\gg 1$ this is natural
assumption. Normalizing the wave function $\Psi({\bf r}, t)$ to unit
probability yields a relation of the form
$$\sum_{m=n-1}^{n+1}A_{m}^{\ast}A_{m}=1.$$
In what follows an explicit form of $A_m$ is immaterial but different
relations between these coefficients are of great importance.
In particular, we need the following equation
$${\frak A}_{\rm KG}=A_{n}^{\ast}A_{n-1}+A_{n+1}^{\ast}A_{n}=\frac{2}{3}.$$

Now let us calculate the expectation value of the momentum operator
$\hat{{\bf P}}$ in the state $\Psi({\bf r},t)$.
In the first step we calculate matrix elements of
$\langle\psi_{m^{\prime}}|\hat{{\bf P}}|\psi_{m}\rangle$ type with
respect to stationary solutions to the Klein-Gordon equation. Then,
using the semiclassical condition ($n\gg 1$) for each matrix element
[see Eqs.\ (\ref{P10})] we write ${\langle\Psi|\hat{{\bf
P}}|\Psi\rangle}_{t}$ in explicit form to give \widetext $${\langle
\hat{P}_{x}\rangle}_{t}=\frac{i}{2}m_0cb_{\perp}\left[ {\frak A}_{\rm
KG}\exp(i\omega t)-{\frak A}^{\ast}_{\rm KG}\exp(-i\omega t) \right]=
-\frac{2}{3}m_0cb_{\perp}\sin\omega t,$$
$${\langle \hat{P}_{y}\rangle}_{t}=\frac{1}{2}m_0cb_{\perp}\left[
{\frak A}_{\rm KG}\exp(i\omega t)-{\frak A}^{\ast}_{\rm KG}\exp(-i\omega t)
\right]=
\frac{2}{3}m_0cb_{\perp}\cos\omega t,$$
$${\langle \hat{P}_{z}\rangle}_{t}=m_0cb_{z}.$$
%\narrowtext
Here
$$\omega=\frac{m_0c^2}{\hbar}(B_{n+1}-B_{n})=
\frac{e_0H}{m_0c^2B_n}\rightarrow\frac{e_0H}{m_0c^2\gamma}$$
is the frequency which, in essence, coincides with the cyclotron
frequency of rotation of a classical particle in a plane perpendicular
to the magnetic field vector. In this approximation,
$B_{n}\rightarrow\gamma$, where $\gamma$ is the Lorentz factor.

This leads us to conclude that the expectation values
differ from the corresponding classical solutions to the equation
of motion of a charged particle only by the factor ($2/3$) in
the ${\langle \hat{P}_{x}\rangle}_{t}$ and
${\langle \hat{P}_{y}\rangle}_{t}$ components. Note that the same result
was also obtained in~\cite{6} for an ordinary one-dimensional quantum
harmonic oscillator. In the discussion below we will show that this factor
can be made as large as unity.

\section{Charge motion and spin precession of a Dirac-Pauli
particle} \label{sec3}

When considering the Dirac-Pauli particle the spin polarization of the
particle must be taken into account.
A superposition of exact solutions to the Dirac-Pauli equation corresponding
to three neighboring energy levels $n-1$, $n$, and $n+1$ has the form
\begin{equation}
\Psi({\bf r},t)=\sum_{\zeta}\sum_{m=n-1}^{n+1}A_{\zeta m}
\psi_{m\zeta}({\bf r})\exp(-\frac{i}{\hbar}m_0c^2B_{m\zeta}t),
\label{1}
\end{equation}
where $\psi_{m}({\bf r})$ are the stationary solutions to the Dirac-Pauli
equation (see Appendix\ \ref{appendixB}) and $A_{\zeta m}$ are expansion
coefficients.
Note that in this case,
allowance is made for polarization states defined by a quantum number
$\zeta=\pm 1$.
Let us assume that a given longitudinal-polarization state at time
$t=0$\footnote{It is known~\cite{5}
that, with regard to the anomalous magnetic moment of an electron, the
operator
$(\bbox{\sigma}\cdot\hat{{\bf P}})$ is not an integral of motion.}
for an electron located at level $m$ is of the form
\begin{equation}
(\bbox{\sigma}\cdot\hat{{\bf P}})\sum_{\zeta}A_{\zeta m}\psi_{m \zeta}=
\lambda_{m}A_{\zeta m}\psi_{m \zeta}.
\label{aaa}
\end{equation}
Having solved Eq.\ (\ref{aaa}) we find the eigenvalue
$$\lambda_{m}=\varepsilon\sqrt{b_{\perp}^2+b_z^2},\quad \varepsilon=\pm 1.$$
In the semiclassical approximation, it can be supposed that for all values
of $m$, the spin coefficients are the same, i.e.,
$C_i(m,\zeta)\simeq C_i(n,\zeta)$ and, moreover, $\lambda_m\simeq\lambda_n$
[see Eqs.\ (\ref{R.5}) and (\ref{R.6})]. Then we have
$$A_{1m}=\kappa A_{-1m},\quad
\kappa=\frac{b_z+\varepsilon b\sqrt{b_{\perp}^2+b_z^2}}
{B_{n\zeta}b_{\perp}}.$$
Let us also require that the wave function $\Psi({\bf r},t)$ be normalized
to unit probability. As a result, we get
\begin{equation}
\sum_{\zeta}\sum_{m=n-1}^{n+1}A_{\zeta m}^{\ast}A_{\zeta m}=1.
\label{2}
\end{equation}
In what follows we need the relations
\begin{mathletters}
\label{3}
\begin{equation}
{\frak A}_{1}=\sum_{m=n-1}^n(A^{\ast}_{1m}A_{1m+1}+
A^{\ast}_{-1m}A_{-1m+1})=\frac{2}{3},
\end{equation}
\begin{eqnarray}
{\frak A}_{2}&=&\sum_{m=n-1}^{n}A^{\ast}_{1m}A_{-1m+1}\nonumber\\
&=&\sum_{m=n-1}^{n}A^{\ast}_{-1m}A_{1m+1}=\frac{2}{3}
\frac{\kappa}{\kappa+1},
\end{eqnarray}
\begin{equation}
{\frak A}_{3}=\sum_{m=n-1}^{n+1}A_{1m}^{\ast}A_{-1m}=
\frac{\kappa}{\kappa^2+1},
\end{equation}
\begin{equation}
{\frak A}_{4}=\sum_{m=n-1}^{n+1}(A_{1m}^{\ast}A_{1m}-
A_{-1m}^{\ast}A_{-1m})=\frac{\kappa^2-1}{\kappa^2+1}.
\end{equation}
\end{mathletters}

To simplify the calculations we suppose that, as an electron makes
a transition from one level
to another, the projection of the momentum of the electron onto the
direction of the magnetic field is conserved.
In addition, random deviations of the orbital center of the electron
due to radiation are neglected. These restrictions imply that
$b^{\prime}_z=b_z$ and $s^{\prime}=s$ (see Appendix\ \ref{appendixB}).

Now, by analogy to a spinless particle, we find the expectation value
of the operator $\hat{{\bf P}}$ in Dirac-Pauli particle state\
(\ref{1}) (see the corresponding matrix
elements in Appendix \ \ref{appendixB}).  As a result of
straightforward but cumbersome calculations, we derive, in view of
these restrictions, the following equations:  \widetext
\begin{mathletters} \label{4} \begin{equation} {\langle
\hat{P}_{x}\rangle}_{t}=\frac{i}{2}m_0cb_{\perp}\left[ {\frak
A}_{1}\exp(i\omega t)-{\frak A}^{\ast}_{1}\exp(-i\omega t)\right]=
-\frac{2}{3}m_0cb_{\perp}\sin\omega t,
\label{4.1}
\end{equation}
\begin{equation}
{\langle \hat{P}_{y}\rangle}_{t}=\frac{1}{2}m_0cb_{\perp}\left[
{\frak A}_{1}\exp(i\omega t)-{\frak A}^{\ast}_{1}\exp(-i\omega t)\right]=
\frac{2}{3}m_0cb_{\perp}\cos\omega t,
\label{4.2}
\end{equation}
\begin{equation}
{\langle P_{z}\rangle}_{t}=m_0cb_{z}.
\end{equation}
\end{mathletters}
%\narrowtext
Here
$$\omega=\frac{m_0c^2}{\hbar}(B_{n+1\zeta}-B_{n\zeta})=
\frac{e_0H}{m_0c^2B_n\zeta}\rightarrow\frac{e_0H}{m_0c^2\gamma}$$
is the frequency which in the BMT approximation (the charge motion
is independent of the spin precession) coincides with
the cyclotron frequency of rotation of an
electron in the plane perpendicular to the magnetic-field vector.
In this approximation, $B_{n\zeta}\rightarrow\gamma$.

A similar procedure applies to calculations of the expectation value
of the spin operator $\hat{S}^{\mu}$ (see~\cite{7}) to yield
$$\hat{S}^{\mu}=\left\lgroup\frac{1}{m_0c}(\bbox{\sigma}\cdot\hat{{\bf P}}),\
\rho_3\bbox{\sigma}+\frac{1}{m_0c}\rho_1\hat{{\bf P}}+
\frac{\mu_a}{m_0c^2}\rho_3{\bf H}\right\rgroup.$$

Taking into account Eqs.\ (\ref{3}), we obtain
\begin{mathletters}
\label{5}
\begin{equation}
{\langle \hat{S}_{x}\rangle}_{t}=-\frac{2}{3}\zeta_{\perp}
(\cos\omega t\sin\Omega_a t+b\sin\omega t\cos\Omega_a t),
\label{5.1}
\end{equation}
\begin{equation}
{\langle \hat{S}_{y}\rangle}_{t}=-\frac{2}{3}\zeta_{\perp}
(\sin\omega t\sin\Omega_a t-b\cos\omega t\cos\Omega_a t),
\label{5.2}
\end{equation}
\begin{equation}
{\langle \hat{S}_{z}\rangle}_{t}=\frac{B_{n\zeta}}{b}\zeta_{z}+
\frac{b_zb_{\perp}}{b}\zeta_{\perp}\cos\Omega_a t,
\end{equation}
\begin{equation}
{\langle \hat{S}^{0}\rangle}_{t}=\frac{b_z}{b}\zeta_{z}+B_{n\zeta}
\frac{b_{\perp}}{b}\zeta_{\perp}\cos\Omega_a t.
\end{equation}
\end{mathletters}
Here $\zeta^2_{z}=1-\zeta^2_{\perp}$,
$\zeta_{\perp}={2\kappa}({\kappa^2+1})^{-1}$ are constants
which, as we will be seen later, are analogous to some constants in
the classical spin theory.
Thus, as in the classical approach, the total precession is determined not
only by the cyclotron frequency $\omega$ but also by the anomalous
frequency
\begin{eqnarray*}
\Omega_a=\frac{m_0c^2}{\hbar}(B_{n\zeta}-B_{n\zeta^{\prime}})=&&
2\frac{\mu_a H}{\hbar}\frac{b}{\sqrt{b_z^2+b^2}}\\
&&\rightarrow
\frac{\alpha}{2\pi}\frac{e_0H}{m_0c}\sqrt{1-\beta_z^2}.
\end{eqnarray*}

Note that in the BMT approximation, the term comprising the anomalous
magnetic moment in the spin operator $\hat{S}^{\mu}$
is immaterial for result\ (\ref{5}).

We see that by analogy with the spinless charged particle,
the results obtained in Eqs.\ (\ref{4.1}), (\ref{4.2}), (\ref{5.1}) and
(\ref{5.2}) are slightly
different from the corresponding solutions to the classical equations of
charge motion and spin precession by the factor ($2/3$).

\section{Semiclassical correspondence principle}
\label{sec4}

To construct a more precise semiclassical theory of charge motion and
spin precession, we will consider a wave packet involving
exact solutions to Dirac-Pauli equation\ (\ref{R.4}) corresponding to
closely spaced energy levels\footnote{It can be shown that there is
a simpler form of this construction for the motion of a spinless
charged particle.}. In this case, wave function\ (\ref{1}) can be
written in the following form:
\begin{equation}
\Psi({\bf r},t)=\sum_{\zeta}\sum_{m\in\Re}A_{\zeta m}
\psi_{m\zeta}({\bf r})\exp(-\frac{i}{\hbar}m_0c^2B_{m\zeta}t),
\label{6}
\end{equation}
where $\Re$ is a set of values of the principle quantum number.
Let us assume that each value of $m$ in Eq.\ (\ref{6}) belonging to the set
is much higher than unity, i.e., $m\gg 1$. At the same time, we have
$$n\gg N=m_{\rm max}-m_{\rm min}\gg 1,$$
where $m_{\rm max}$ and $m_{\rm min}$ are maximum and minimum values
belonging to $\Re$, $n$ is the quantum number from $\Re$ defining some
energy level to which a given classical trajectory corresponds.
Let us further assume that the other energy levels from the set are
symmetric about level $n$.

The next natural step in our method is to derive the coefficients
$A_{\zeta m}$. The central idea
of the derivation is the same as in Eq.\ (\ref{2}), i.e.,
$$A_{1m}=\kappa
A_{-1m},\quad \sum_{\zeta}\sum_{m\in\Re}A_{\zeta m}^{\ast}A_{\zeta
m}=1.$$
Then, relations\ (\ref{3}) can be easily extended to the case under
consideration, namely,
\begin{mathletters}
\label{aaa1}
\begin{equation}
{\frak A}_{1}=\sum_{m=m_{\rm min}}^{m_{\rm
max}-1}(A^{\ast}_{1m}A_{1m+1}+
A^{\ast}_{-1m}A_{-1m+1})=\frac{N-1}{N},
\end{equation}
\begin{eqnarray}
{\frak A}_{2}&=&\sum_{m=m_{\rm min}}^{m_{\rm max}-1}A^{\ast}_{1m}A_{-1m+1}
\nonumber\\
&=&\sum_{m=m_{\rm min}}^{m_{\rm max}-1}A^{\ast}_{-1m}A_{1m+1}=\frac{N-1}{N}
\frac{\kappa}{\kappa+1},
\end{eqnarray}
\begin{equation}
{\frak A}_{3}=\sum_{m\in\Re}A_{1m}^{\ast}A_{-1m}=
\frac{\kappa}{\kappa^2+1},
\end{equation}
\begin{equation}
{\frak A}_{4}=\sum_{m\in\Re}(A_{1m}^{\ast}A_{1m}-
A_{-1m}^{\ast}A_{-1m})=\frac{\kappa^2-1}{\kappa^2+1}.
\end{equation}
\end{mathletters}

Following the above procedure of calculating the expectation values
of the momentum and spin operators and using Eqs.\ (\ref{aaa1}) we obtain
$${\langle \hat{P}_{x}\rangle}_{t}=-\frac{N-1}{N}m_0cb_{\perp}\sin\omega t,$$
$${\langle \hat{P}_{y}\rangle}_{t}=
\frac{N-1}{N}m_0cb_{\perp}\cos\omega t,$$
$${\langle \hat{P}_{z}\rangle}_{t}=m_0cb_{z},$$
$${\langle \hat{S}_{x}\rangle}_{t}=-\frac{N-1}{N}\zeta_{\perp}
(\cos\omega t\sin\Omega_a t+b\sin\omega t\cos\Omega_a t),$$
$${\langle \hat{S}_{y}\rangle}_{t}=-\frac{N-1}{N}\zeta_{\perp}
(\sin\omega t\sin\Omega_a t-b\cos\omega t\cos\Omega_a t),$$
$${\langle \hat{S}_{z}\rangle}_{t}=\frac{B_{n\zeta}}{b}\zeta_{z}+
\frac{b_zb_{\perp}}{b}\zeta_{\perp}\cos\Omega_a t,$$
$${\langle \hat{S}^{0}\rangle}_{t}=\frac{b_z}{b}\zeta_{z}+
B_{n\zeta}\frac{b_{\perp}}{b}\zeta_{\perp}\cos\Omega_a t.$$

Thus, for $N\gg 1$, the difference of  ${\langle\hat{P}_x\rangle}_t$,
${\langle\hat{P}_y\rangle}_t$, ${\langle\hat{S}_x\rangle}_t$, and
${\langle\hat{S}_y\rangle}_t$ from the results obtained by classical
theory is eliminated, and the time evolution of
the expectation values ${\langle\hat{{\bf P}}\rangle}_t$ and ${\langle
\hat{S}^{\mu}\rangle}_t$ are made to coincide with the solutions to the
corresponding classical equations~\cite{8}.

It is easy to see that the anomalous magnetic moment
of an electron has effect not only on the time behavior of spin projection
onto the direction of motion (longitudinal polarization)
\cite{5} but
also on the behavior of spin in the plane perpendicular to the
magnetic-field vector, i.e., on the time evolution of
${\langle\hat{S}_x\rangle}_t$ and ${\langle \hat{S}_y\rangle}_t$.

\section{Concluding remarks}

It should be noted that ${\langle\hat{S}^{\mu}\rangle}_t$ derived in
Sec.\ \ref{sec4} satisfy the relations given by the
classical spin theory:  $${\langle \hat{S}^{\mu}\rangle}_t{\langle
\hat{P}_{\mu}\rangle}_t=0, \quad {\langle
\hat{S}^{\mu}\rangle}_t{\langle \hat{S}_{\mu}\rangle}_t=1.$$ Moreover,
the procedure under review can be extended to the tensor
operator\cite{7}
$$\hat{\Pi}^{\mu\nu}=\left\lgroup\hat{\bbox{\Phi}},\hat{\bbox{\Pi}}
\right\rgroup,$$
$$\hat{\bbox{\Phi}}=-\frac{1}{m_0c}\rho_3(\bbox{\sigma}
\times\hat{{\bf P}})+
\frac{\mu_a}{m_0c^2}\rho_1(\bbox{\sigma}\times{\bf H}),$$
$$\hat{\bbox{\Pi}}=\bbox{\sigma}+\frac{1}{m_0c}\rho_2(\bbox{\sigma}\times
\hat{{\bf P}})+\frac{\mu_a}{m_0c^2}{\bf H},$$
which, in the BMT approximation, is related to $\hat{S}^{\mu}$ by
the formula~\cite{1}:
$$\hat{\Pi}^{\mu\nu}=\frac{1}{m_0c}\varepsilon^{\mu\nu\alpha\beta}
\hat{S}_{\alpha}\hat{P}_{\beta}.$$

Thus, the expectation values ${\langle \hat{S}^{\mu}\rangle}_t$
and ${\langle \hat{\bf P}\rangle}_t$ as well as
${\langle \hat{\Pi}^{\mu\nu}\rangle}_t$ calculated following the
procedure discussed in Sec.\ \ref{sec4} obey the classical equation of
charge motion and spin precession. The approach introduced in this work
substantiates the use of the semiclassical method for gaining an insight
into physical phenomena associated with high-energy charged particles.
In particular, the semiclassical approach can be the basis for a
classical model for certain purely quantum processes. For example, it
can be used to establish a relationship between spin-flip transitions
and classical spin precession \cite{2}.

\acknowledgments

The authors would like to thank Professor V.G. Bagrov for support and
valuable discussion.

\appendix
\section{}
\label{appendixA}

Motion of a scalar particle in a constant and homogeneous magnetic field
${\bf H}=(0,0,H)$ is given by the Klein-Gordon equation
$$(\frac{1}{c^2}\hat{E}^2-\hat{{\bf P}}^2+m_0^2c^2)\psi({\bf r},t)=0,$$
where $\hat{E}=i\hbar{\partial}/{\partial t}$ is the energy operator,
$\hat{{\bf P}}=\hat{{\bf p}}-(e/c){\bf A}$ is the kinetic momentum
operator,
${\bf A}=(-Hy/2,Hx/2,0)$, $e=-e_0<0$ is the charge, and
$m_0$ is the rest mass of the particle.

It is known that in this case, the wave function $\psi({\bf r},t)$
is to be the eigenfunction of the energy operator $\hat{E}$
as well as $z$-components of the kinetic momentum operator
$\hat{\bf P}$ and orbital angular momentum operator
$\hat{\bf L}={\bf r}\times\hat{\bf p}$
onto the direction of magnetic field vector, i.e.,
$$\hat{E}\psi=E_n\psi,\quad \hat{P}_z\psi=p_z\psi,
\quad \hat{L}_z\psi=\hbar l\psi.$$
Then, in terms of cylindrical coordinates ($r$, $\varphi$, $z$),
which are the most suitable ones in this case, the solution to the
Klein-Gordon equation has the form~\cite{10}
\begin{equation}
\psi({\bf r},t)=\frac{\sqrt{e_0H}}{\sqrt{L}\sqrt{2\pi\hbar c}}
e^{-{i}E_nt/\hbar}e^{{i}p_zz/\hbar}e^{il\varphi}I_{n,s}(\rho).
\label{kg1}
\end{equation}
The function $I_{n,s}(\rho)$ in Eq.\ (\ref{kg1}) is defined by
the Laguerre polynomials $Q^{l}_{s}(\rho)$ with help of the following
relation
$$I_{n,s}(\rho)=\frac{1}{\sqrt{n!s!}}e^{-\frac{1}{2}\rho}Q^{n-s}_{s}
\rho^{\frac{n-s}{2}},\quad\rho=\frac{e_0H}{2\hbar c}r^2.$$
Here $n=l+s=0,1,2,\ldots$ is the principle quantum number,
$s=0,1,2,\ldots$ and $l=0,\pm1,\pm2,\ldots$ are radial and azimuthal
quantum numbers.

The energy of a particle is written as
$$E_n=m_0c^2B_n=m_0c^2\sqrt{1+b_z^2+b_{\perp}^2},\quad
b_{\perp}=2\sqrt{\frac{\mu_0H}{m_0c^2}(n+\frac{1}{2})},$$
where $b_z={p_z}/{m_0c}$ is the projection of the momentum onto the
direction of the field and $\mu_0={e_0\hbar}/{2m_0c}$ is the Bohr magneton.

Matrix elements of the momentum operator of scalar particle have the form
\begin{mathletters}
\label{P10}
\begin{equation}
\langle\psi_{m^{\prime}}|\hat{P}_x|\psi_{m}\rangle=
\frac{i}{2}m_0cb_{\perp}(\delta_{m^{\prime}-m,+1}-\delta_{m^{\prime}-m,-1}),
\end{equation}
\begin{equation}
\langle\psi_{m^{\prime}}|\hat{P}_y|\psi_{m}\rangle=
\frac{1}{2}m_0cb_{\perp}(\delta_{m^{\prime}-m,+1}+\delta_{m^{\prime}-m,-1}),
\end{equation}
\begin{equation}
\langle\psi_{m^{\prime}}|\hat{P}_z|\psi_{m}\rangle=
m_0cb_{z}\delta_{m^{\prime}m}.
\end{equation}
\end{mathletters}

\section{}
\label{appendixB}

In the paper we use the solutions to the Dirac-Pauli equation
$$i\hbar\frac{\partial\psi}{\partial t}=\hat{H}\psi,\quad
\hat{H}=c(\bbox{\alpha}\cdot\hat{{\bf P}})+\rho_3m_0c^2+
\mu_a\rho_3(\bbox{\sigma}\cdot{\bf H}),$$
where $\hat{{\bf P}}=\hat{{\bf p}}-(e/c){\bf A}$ is the kinetic momentum
operator, ${\bf A}=(-Hy/2,Hx/2,0)$, ${\bf H}$ is the external magnetic field,
$\mu_a=({\alpha}/{2\pi})\mu_0$ is the
anomalous part of electron magnetic moment, $e=-e_0<0$ and $m_0$ are
the charge and rest mass of electron.

It is known~\cite{5} that in this case complete set of commuting operators
characterizing the quantum state of the particle consists of the Hamiltonian
$\hat{H}$, the projections of momentum operator $\hat{\bf P}$ and
total angular momentum operator $\hat{\bf J}=\hat{\bf L}+\frac{1}{2}\hbar
\bbox{\sigma}$ onto the direction of magnetic field
vector which, for definiteness, is oriented along the $Z$-axis,
and polarization operator. According to~\cite{5}, in a homogeneous
magnetic field the operator
$$\hat{\Pi}_{z}=\sigma_z+\frac{1}{m_0c}{\rho_2(\bbox{\sigma}\times
\hat{{\bf P}})_z}+\frac{\mu_aH}{c}$$
is taken as spin integral of motion characterizing
the polarization of electron spin relative to the magnetic field vector.
Then we have
\begin{mathletters}
\label{R.3}
\begin{equation}
\hat{H}\psi=E_{n\zeta}\psi,\quad\hat{P}_z\psi=p_z\psi,\quad
\hat{J}_z\psi=\hbar(l-\frac{1}{2})\psi,
\end{equation}
\begin{equation}
\hat{\Pi}_z\psi=\zeta b\psi,\quad\zeta=\pm 1.
\end{equation}
\end{mathletters}

In terms of cylindrical coordinates ($r$, $\varphi$, $z$),
the wave function being a solution to the Dirac-Pauli equation and
satisfying Eqs.\ (\ref{R.3}) has the form~\cite{5}
\begin{equation}
\psi({\bf r},t)=\frac{\sqrt{e_0H}}{\sqrt{L}\sqrt{2\pi\hbar c}}
e^{-{i}E_{n\zeta}t/\hbar}e^{{i}p_zz/\hbar}
e^{i(l-1)\varphi}f(\rho, \varphi),
\label{R.4}
\end{equation}
where
$$f(\rho, \varphi)=\left\lgroup
\begin{array}{l}
C_{1}I_{n-1,s}(\rho) \\
iC_{2}I_{n,s}(\rho)e^{i\varphi} \\
C_{3}I_{n-1,s}(\rho) \\
iC_{4}I_{n,s}(\rho)e^{i\varphi}
\end{array}
\right\rgroup.$$
Here $n=l+s=0,1,2,\ldots$ is the principle quantum number,
$s=0,1,2,\ldots$ is the radial quantum number, and $l=0,\pm1,\pm2,\ldots$ is
the azimuthal quantum number.
In dimensionless form, the spin coefficients $C_i$ are determined as
\begin{eqnarray}
C_1&=&\frac{\zeta}{2}\sqrt{\frac{1}{2}(1+\zeta\frac{1}{b})}
\left[\sqrt{1+\frac{b_z}{B_{n\zeta}}}+\zeta\sqrt{1-\frac{b_z}{B_{n\zeta}}}
\right], \nonumber \\
C_2&=&\frac{\zeta}{2}\sqrt{\frac{1}{2}(1-\zeta\frac{1}{b})}
\left[\sqrt{1-\frac{b_z}{B_{n\zeta}}}-\zeta\sqrt{1+\frac{b_z}{B_{n\zeta}}}
\right], \nonumber \\
C_3&=&\frac{\zeta}{2}\sqrt{\frac{1}{2}(1+\zeta\frac{1}{b})}
\left[\sqrt{1+\frac{b_z}{B_{n\zeta}}}-\zeta\sqrt{1-\frac{b_z}{B_{n\zeta}}}
\right], \nonumber \\
C_4&=&\frac{1}{2}\sqrt{\frac{1}{2}(1-\zeta\frac{1}{b})}
\left[\sqrt{1+\frac{b_z}{B_{n\zeta}}}+\zeta\sqrt{1-\frac{b_z}{B_{n\zeta}}}
\right],
\label{R.5}
\end{eqnarray}
where the energy of electron is given by the expression
\begin{equation}
B_{n\zeta}=\frac{E_{n\zeta}}{m_0c^2}
=\sqrt{b_z^2+{\left(\sqrt{1+b_{\perp}^2}+
\zeta\frac{\mu_aH}{m_0c^2}\right)}^2},
\label{R.6}
\end{equation}
$$b_z=\frac{p_z}{m_0c},\quad b_{\perp}=2\sqrt{\frac{\mu_0H}{m_0c^2}n},
\quad b=\sqrt{1+b_{\perp}^2}.$$

Let us consider exact forms of all necessary matrix elements of the
momentum and spin operators.
$$\langle\psi_{m^{\prime}\zeta^{\prime}}|\hat{P}_x|\psi_{m\zeta}\rangle=
\frac{i}{2}m_0cb_{\perp}(\delta_{m^{\prime}-m,+1}-\delta_{m^{\prime}-m,-1})
\delta_{\zeta^{\prime}\zeta},$$
$$\langle\psi_{m^{\prime}\zeta^{\prime}}|\hat{P}_y|\psi_{m\zeta}\rangle=
\frac{1}{2}m_0cb_{\perp}(\delta_{m^{\prime}-m,+1}+\delta_{m^{\prime}-m,-1})
\delta_{\zeta^{\prime}\zeta},$$
$$\langle\psi_{m^{\prime}\zeta^{\prime}}|\hat{P}_z|\psi_{m\zeta}\rangle=
m_0cb_{z}\delta_{m^{\prime}m}\delta_{\zeta^{\prime}\zeta},$$
\widetext
$$\langle\psi_{m^{\prime}\zeta^{\prime}}|\hat{S}_x|\psi_{m\zeta}\rangle=
\frac{i}{2}(b-\zeta(\delta_{m^{\prime}-m,+1}-\delta_{m^{\prime}-m,-1}))
(\delta_{m^{\prime}-m,+1}-\delta_{m^{\prime}-m,-1})
\delta_{-\zeta^{\prime}\zeta},$$
$$\langle\psi_{m^{\prime}\zeta^{\prime}}|\hat{S}_y|\psi_{m\zeta}\rangle=
\frac{1}{2}(b-\zeta(\delta_{m^{\prime}-m,+1}-\delta_{m^{\prime}-m,-1}))
(\delta_{m^{\prime}-m,+1}+\delta_{m^{\prime}-m,-1})
\delta_{-\zeta^{\prime}\zeta},$$
%\narrowtext
$$\langle\psi_{m^{\prime}\zeta^{\prime}}|\hat{S}_z|\psi_{m\zeta}\rangle=
(\zeta\frac{B_{n\zeta}}{b}\delta_{\zeta^{\prime}\zeta}+\frac{b_{\perp}b_z}{b}
\delta_{-\zeta^{\prime}\zeta})\delta_{m^{\prime}m},$$
$$\langle\psi_{m^{\prime}\zeta^{\prime}}|\hat{S}^0|\psi_{m\zeta}\rangle=
(\zeta\frac{b_{z}}{b}\delta_{\zeta^{\prime}\zeta}+
\frac{B_{n\zeta}b_{\perp}}{b}
\delta_{-\zeta^{\prime}\zeta})\delta_{m^{\prime}m}.$$

\end{document}